**Towards chemically neutral carbon cleaning processes: Plasma cleaning of Ni, Rh, and Al reflective optical coatings and thin Al filters for Free Electron Lasers and synchrotron beamline applications.**

H. Moreno Fernández[a*], M. Zangrando[b,c], G. Sauthier[d], A.R. Goñi[e,f], V. Carlino[g], and E. Pellegrin[a*]

[a]CELLS-ALBA, Carrer de la Llum 2-26, E-08290 Cerdanyola del Vallès, Spain

[b]Elettra-Sincrotrone Trieste, Strada Statale 14 km 163-5, 34149 Basovizza, Trieste, Italy

[c]IOM-CNR, Strada Statale 14 km 163-5, 34149 Basovizza, Trieste, Italy

[d]ICN2, UAB Campus, E-08193, Bellaterra, Spain

[e]Institut de Ciència de Materials de Barcelona (ICMAB-CSIC), Campus UAB, Bellaterra, Spain

[f]ICREA, Passeig Lluís Companys 23, 08010 Barcelona, Spain

[g]ibss Group Inc., Burlingame, CA 94010, USA

## ABSTRACT

The choice of a reflective optical coating or filter material has to be adapted to the intended field of application. This is mainly determined by the required photon energy range or by the required reflection angle. Among various materials, nickel and rhodium are standard materials used as reflective coatings for synchrotron mirrors. Conversely, Aluminum is one of the most commonly used materials for extreme ultraviolet (EUV) and soft X-ray filters. However, both of these types of optics are subject to carbon contamination, being increasingly problematic for the operation of the high-performance free electron laser (FEL) and synchrotron beamlines. For this reason, an inductively coupled plasma (ICP) source has been used in conjunction with $N_2/O_2/H_2$ and $N_2/H_2$ feedstock gas plasmas. Results from the chemical surface analysis of the above materials before and after plasma treatment using X-ray photoelectron spectroscopy (XPS) are reported. We conclude that a favorable combination of an $N_2/H_2$ plasma feedstock gas mixture leads to the best chemical surface preservation of Ni, Rh, and Al while removing the carbon contaminations. However, this feedstock gas mixture does not remove C contaminations as rapidly as, e.g., a $N_2/O_2/H_2$ plasma which induces the surface formation of NiO and NiOOH in Ni and RhOOH in Rh foils. As an applied case, we demonstrate the successful carbon removal from ultrathin Al filters previously used at the FERMI FEL1 using a $N_2/H_2$ plasma.

## 1 INTRODUCTION

As of today, synchrotron radiation is an established powerful tool for a broad range of research in science and technology. The increase of synchrotron laboratories worldwide highlights the interest by commercial, educational, and medical scientific research for obtaining analytical results that are only achievable using accelerator-based high brilliance light sources. New generations of synchrotron sources with enhanced performance such as, e.g., Free Electron Lasers (FELs), offer new possibilities for fundamental and applied research in the field of time-resolved and/or coherence-based experiments. In this context, one pressing practical requirement is to maintain the enhanced performance of these facilities, which includes maintaining reflective and transmission beamline optics as close as possible to their original/pristine state.

Rh, Ni, and Al among others, are the most common optical coatings for synchrotron mirrors, due to their reflectance characteristics, where the material choice depends on the intended field of application and becomes an integral part of the beamline optics design. Aluminum is the most commonly used material for EUV and soft X-Ray filters due to its specific mechanical properties, thermal conductance, and wide photon energy bandpass while blocking – among others - visible light.

However, reflective optics, as well as transmission filters, are subjected to carbon contamination buildup, where the latter represents an increasingly serious problem for the operation of high-performance FEL beamlines. Previous publications have shown that it is possible to clean those carbon contaminations[1-9] by using a different kind of plasma such as, e.g., capacitively coupled plasma (CCP), inductively coupled plasma (ICP), and microwave plasma obtaining different results depending on the gas mixture, RF input power, frequencies and surfaces to be cleaned. In this vein and following previous studies[1-3], the ICP-type ibss model GV10x low-pressure plasma source together with $N_2/O_2/H_2$ and $N_2/H_2$ feedstock gases has been used to clean Ni, Rh, and Al foil materials and ultrathin Al filters for Extreme Ultra-Violet (EUV) applications.

Using X-ray photoelectron spectroscopy (XPS), we report on changes regarding the surface chemistry of the various above metals before and after the different plasma treatments. We conclude that $N_2/H_2$ is a favorable feedstock gas combination for a plasma that leads to the best results in terms of preserving the original surface chemistry for Ni, Rh, and Al metals, especially in comparison with previous studies[1, 2]. Carbon contaminations could still be removed successfully, but not as fast as with $N_2/O_2/H_2$ plasma which, on the downside, induces the formation of NiO and NiOOH in Ni and RhOOH in Rh foils. In addition, results from the successful carbon cleaning of ultrathin Al EUV filters used at the FERMI FEL1[10] is presented.

## 2 EXPERIMENTAL

### *2.1 Materials and cleaning test setup*

A set of Ni, Rh, and Al metal foils with micrometric thickness were contaminated with 50 nm of amorphous carbon (a-C) for simulating a photon-beam induced carbon contamination (Figure 1). On the other hand, self-sustained ultrathin Al filters (Luxel Corporation, Friday Harbor, WA, USA) with 100 nm thickness were contaminated by exposure to the FEL photon beam operation at the FERMI FEL1. All samples were plasma cleaned in an ultra-high vacuum (UHV) chamber (Figure 2), where a 300 l/s turbo-molecular pumping unit was installed in order to sustain a base pressure of $1.9\text{x}10^{-8}$ mbar. The plasma operating pressure was kept at 0.005 mbar (3.75 mTorr) for both $N_2(94\%)/O_2(4\%)/H_2(2\%)$ and $N_2(95\%)/H_2(5\%)$ feedstock gas mixtures supplied via commercial pre-mix gas bottles (Air Liquide). Both gases will be referred to as $N_2/O_2/H_2$ and $N_2/H_2$ from now on. The plasma feedstock gas pressure was kept constant at 0.005 mbar, which corresponds to best cleaning rates that can be obtained when cleaning carbon contaminations on optical components in our setup with the above feedstock gasses. The RF power of the ICP plasma source was maintained at 100 W.

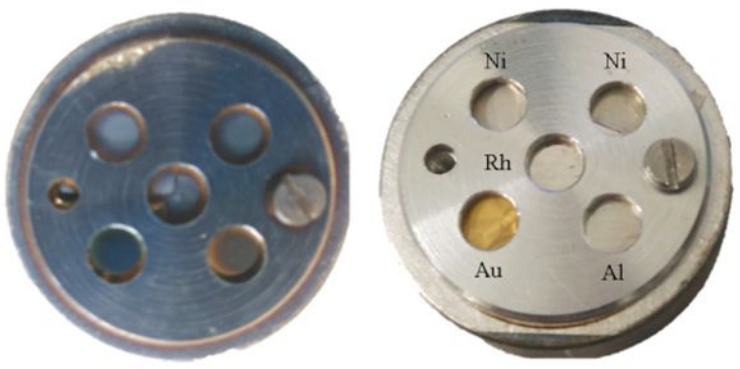

Figure 1: Ni, Rh, Al, and Au metal foils of micrometric thickness installed on a common foil sample holder for plasma treatment. Left: Metal foil sample holder before plasma cleaning, including a 50 nm a-C coating. Right: Metal foil sample holder after plasma cleaning.

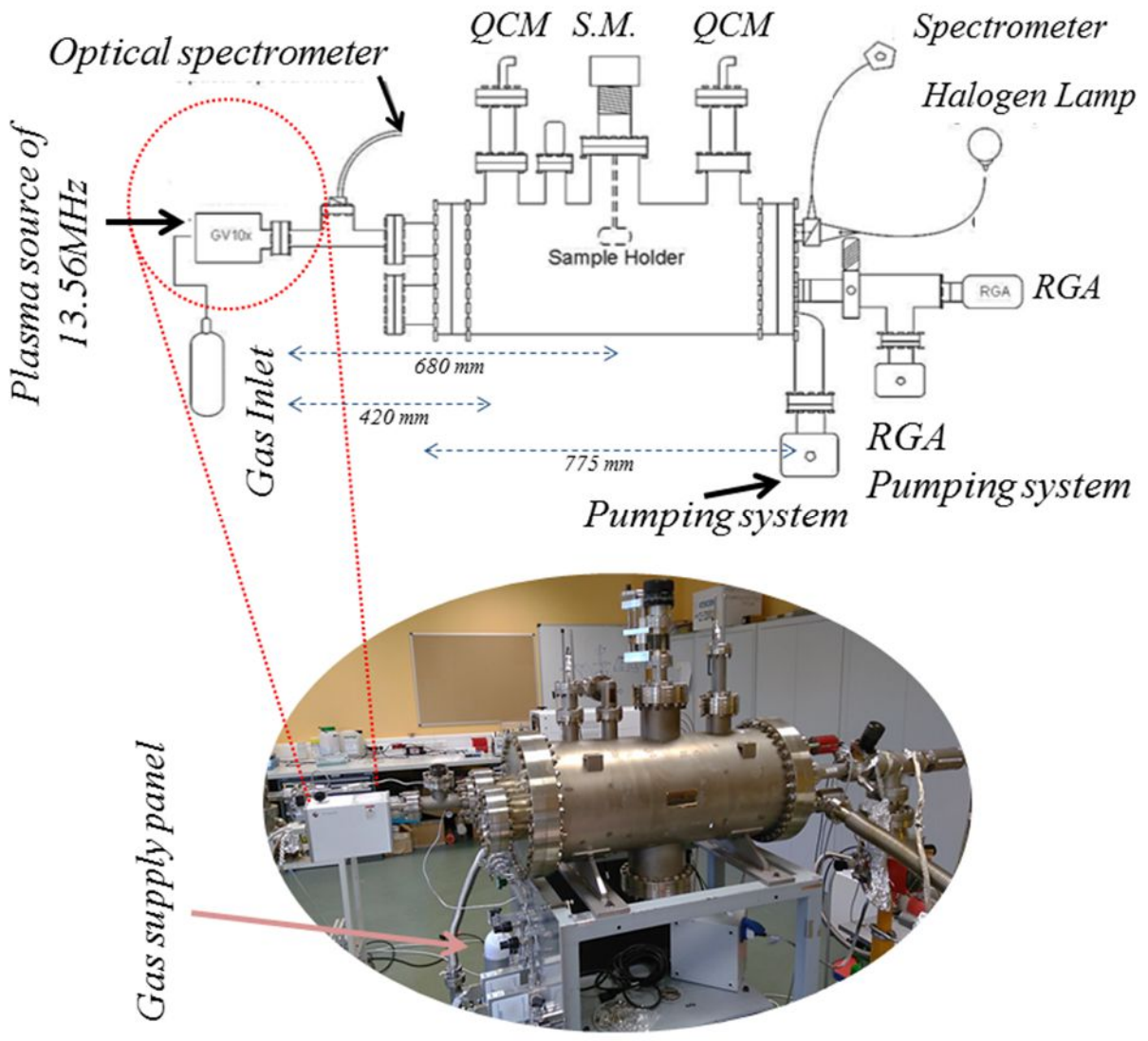

Figure 2: Schematical description of the plasma cleaning chamber (QCM: Quartz crystal monitor; O.F.: Optical fiber; RGA: Residual gas analyzer).

## 2.2 *Plasma source*

The plasma source consists of a commercial ICP source (ibss model GV10x Downstream Asher, Burlingame, CA, USA) operating at an RF frequency of 13.56 MHz. The Downstream Asher operation principle allows for generating the plasma in a separate "remote plasma" volume upstream the UHV chamber including the plasma diagnostics and the optical objects to be cleaned. With the plasma source being localized outside the main chamber, the ionization of the different gasses will only take place within the upstream plasma source. The a-C cleaning rate – as measured by the quartz balance closer to the plasma source (Figure 2) - was 7.2 Å/min and 2.5 Å/min for $N_2/O_2/H_2$ and $N_2/H_2$ plasma, respectively. This significant difference in terms of carbon cleaning rate is related to the small oxygen percentage in the former gas mixture. Differences in terms of cleaning rate were already observed in previous studies[1-3].

## 2.3 *Characterization techniques*

For detecting ionic species generated by the plasma, a differentially pumped residual gas analyzer (RGA) was installed at the remote end of the UHV chamber and an optical spectrometer was connected close to the GV10x plasma source exhaust to check the optical emission spectrum from the downstream plasma.

Plasma-induced changes regarding the sample surface chemistry were analyzed by XPS using a SPECS Phoibos 150 electron energy analyzer (operated at 25 eV pass energy) in conjunction with a monochromatized Al Kα X-ray source (0.65 eV line widths at 1486.6 eV) for samples cleaned with $N_2/O_2/H_2$ plasma. On the other hand, a non-monochromatized Al Kα x-ray source has been used for the analysis of samples cleaned with $N_2/H_2$ plasma.

In addition, in order to contrast the chemical composition and allotropism of the carbon contaminations coming on the Al EUV filters with respect to the deposited a-C, Raman spectroscopy was used to characterize these carbon contaminations using red laser light as an excitation source.

# 3 RESULTS AND DISCUSSION

## *3.1 Nickel Spectra*

Figure 3 shows the Ni 2p XPS spectrum for plasma-cleaned nickel foils using two different plasma as compared with the pristine reference nickel foil (i.e., without any treatment). The binding energy (BE) values ($Ni2p_{3/2}$ and O1s) as obtained from the different plasma treatments are reported in Table I. After the cleaning process using $N_2/H_2$ plasma, no changes were observed regarding the Ni chemical surface state in both the native oxide surface layer as well as the Ni metal foil bulk, unlike expected from previous studies[2], where hydrogen-based $Ar/H_2$-plasma had resulted into a significant reduction of the native $Ni_2O_3$ surface layer of the Ni foil.

In addition, as the intensity ratio between the Ni $2p_{3/2}$ lines for both Ni in $Ni_2O_3$ and in Ni metal remains unchanged, the thickness of the former native oxide layer appears to be unchanged. Regarding the O1s XPS core level data, the peak corresponding to $Ni_2O_3$ for the plasma treated foil appears to be not as well-defined as in the case of the Ni reference foil, most probably due to the non-monochromatized Al Kα source used in the former case.

However, when using the $N_2/O_2/H_2$ plasma the $Ni2p_{3/2}$ spectrum shows a significant change towards a more complex structure with intense satellite lines at higher binding energies corresponding to the $\underline{c}d^8$ and $\underline{c}d^9\underline{L}$ plus the $\underline{c}d^{10}\underline{L}^2$ XPS final states, respectively. The $Ni2p_{3/2}$ line gives two different species at binding energies (BEs) of 853.58 and 855.71 eV as compared with the Ni reference foil, apparently induced by the $N_2/O_2/H_2$ plasma treatment now including a small oxygen concentration. On the basis of their BEs, we assign these peaks to NiO and NiOOH, respectively, which is in agreement with previous studies[11, 12]. The O1s XPS spectrum gives four different peaks for all samples, although the spectrum from $N_2/O_2/H_2$ plasma-treated foil shows appreciable variations. Taking into account these changes, these O1s spectral features can be assigned to different species and the associated BE positions are as follows: 529.06, 530.68, and 531.68 eV for NiO, C-C, and C-O respectively.

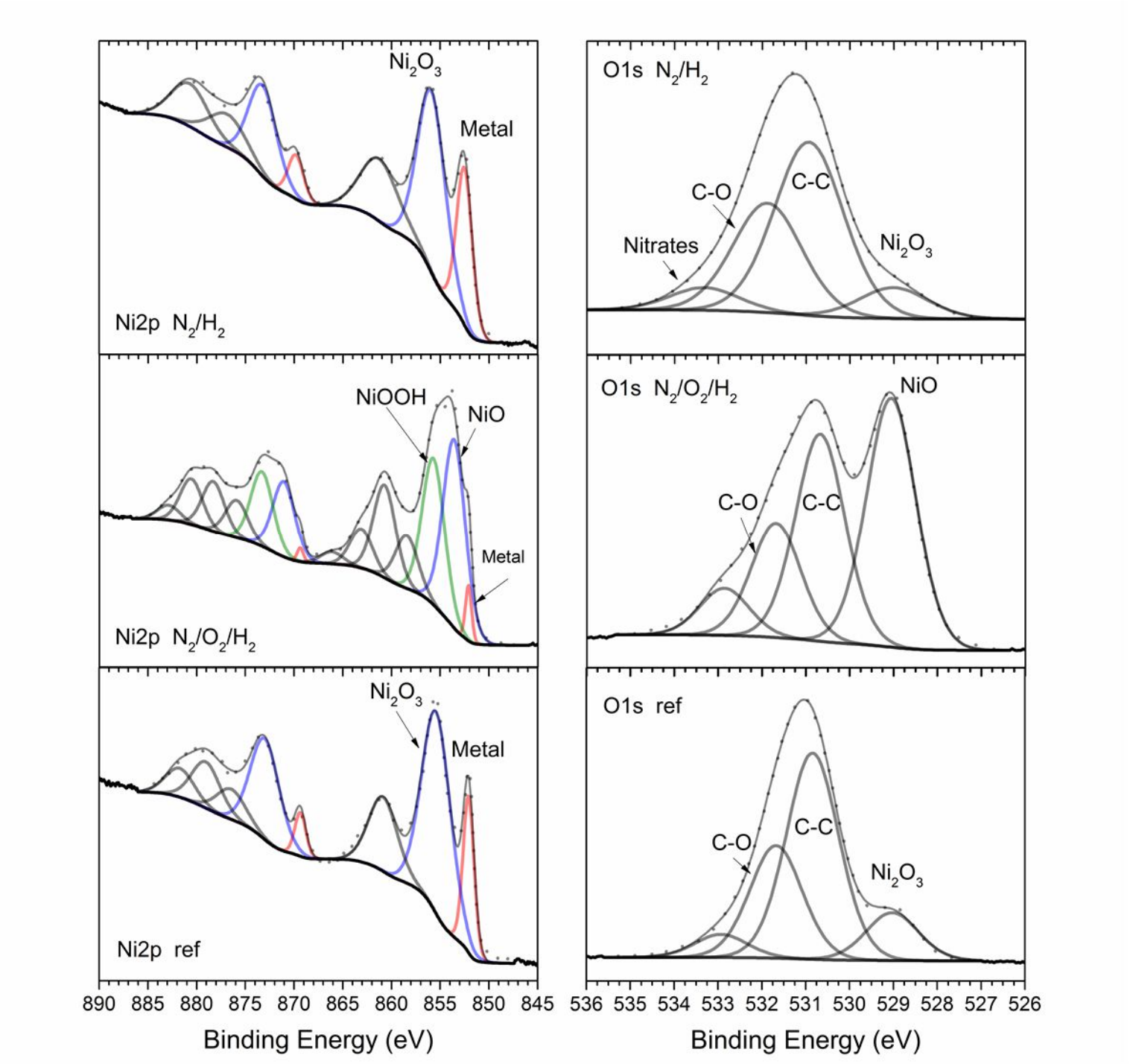

Figure 3: Ni 2p (left column) and O 1s (right column) XPS core level spectra of Ni metal foils. Bottom row: Pristine Ni reference sample; Center row: Ni foil after a-C contamination and subsequent $N_2/O_2/H_2$ plasma treatment; Top row: Ni foil after a-C contamination and subsequent $N_2/H_2$ plasma treatment.

## *3.1 Rhodium Spectra*

Figure 4 shows the Rh 3d XPS spectra for the a-C contaminated and subsequently plasma-cleaned rhodium metal foils with two different plasmas as compared with a pristine reference rhodium foil (i.e., without any treatment). In the case of rhodium, we observe a similar behavior as in the case of nickel after $N_2/H_2$ plasma treatment, where chemical changes could not be observed in contrast to prior expectations based on previous studies. The peak position attributed to $Rh_2O_3$ oxide is shifted by 0.67 eV to higher BE in comparison with the $Rh_2O_3$ oxide from the reference foil (due to the lack of spectral features that would allow for a better fitting).

The reference Rh 3d spectra were fitted with two main peaks at binding energies of Rh $3d_{5/2}$ at 307.04 and 307.6 eV for metallic Rh and $Rh_2O_3$, respectively. The BEs of Rh metal and $Rh_2O_3$ oxide after the $N_2/H_2$ plasma treatment are in fair agreement with the values obtained by Kibis et al.[13] for metallic $Rh^0$ and $Rh^{3+}$ in $Rh_2O_3$ respectively, where peak positions at 307.3 and 308.2 eV, respectively, were obtained. As in the case of the $N_2/H_2$ plasma-treated Ni foil (Figure 3, top row), the $N_2/H_2$ plasma-treated Rh foil exhibits slightly broader spectral lines due to the use of a non-monochromatized Al Kα source.

After the $N_2/O_2/H_2$ plasma treatment, a clearly enhanced oxidation of the Rh foil can be observed from the increase of the $Rh_2O_3$-related $Rh_{5/2}$ peak at 307.93 eV BE (Figure 4). The O1s spectrum now shows four features: The peak formerly observed on the Rh reference foil as well as the $N_2/H_2$ plasma-treated Rh foil at 529.68 eV, plus additional peaks at 531.32, 532.49 and 534.35 eV BE. We attribute those additional peaks to $Rh_2O_3$, rhodium hydroxide groups,[13-15] carbonates and adsorbed $H_2O$. In this context, we associate the Rh $3d_{5/2}$ peak at 309.59 eV with rhodium hydroxide group.

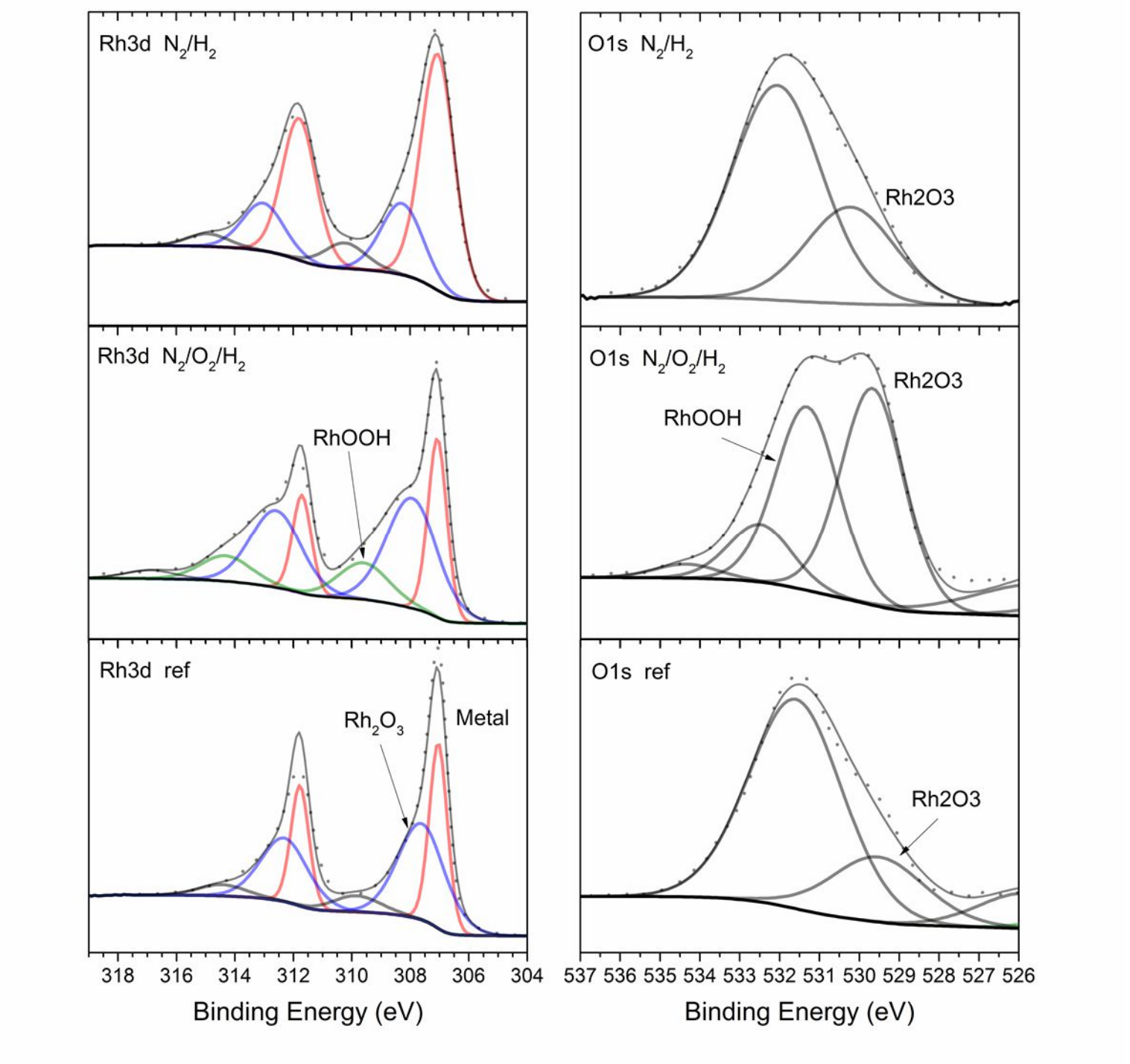

Figure 4: Rh 3d (left column) and O 1s (left column) XPS core level spectra of Rh metal foils. Bottom row: Pristine Rh metal reference sample; Center row: Rh foil after a-C contamination and subsequent $N_2/O_2/H_2$ plasma treatment; Top row: Rh foil after a-C contamination and subsequent $N_2/H_2$ plasma treatment.

At variance with the previous study by Kibis et al.,[13] where $RhO_2$ peaks were found after plasma sputtering and characterized in terms of an additional O1s line at 530.6 eV BE, we could not detect such a feature at this BE position in our O1s XPS spectra. Nevertheless, the XPS data on $RhO_2$ are a bit contradictory due to the fact that the $Rh^{4+}O_2$ in the $Rh_{5/2}$ core-level line has been reported to be in the range of 309-310 eV, which in the case of our $N_2/O_2/H_2$ plasma treatment corresponds to a peak position at 309.59 eV. In that sense, we attribute the O1s line at 531.32 eV and the $Rh3d_{5/2}$ line at 309.59 eV as $Rh^{4+}/Rh^{3+}$-based rhodium oxyhydroxide species.

### *3.1 Aluminum Spectra*

Taking into account the results from the previous cleaning processes, where – besides the expected significant a-C contamination removal – only the $N_2/H_2$ plasma appears to leave the metal surface chemistry unchanged, we thus only performed a $N_2/H_2$ plasma cleaning process on the Al foils. In this way, we avoid a further oxidation of the Al foil surface beyond the native $Al_2O_3$ layer, together with a possible mechanical destabilization via the formation of that brittle oxide. The detailed Al 2p and O1s core level spectra are shown in Figure 5, for Al metal foils before and after a-C contamination plus subsequent $N_2/H_2$ plasma treatment. In both cases, the metal peaks corresponding to the Al $2p_{3/2}$ and $2p_{1/2}$ core level lines can be seen at 72.53 eV and 73.19 eV BE, respectively, while the peak at 75.47 eV is related to the $Al_2O_3$ surface layer.

These Al2p spectral features are in good agreement with values found in the standard web-based XPS literature[16] for Al oxide on Al foil. However, in contrast to values reported in the literature,[17, 18] none of our XPS data on oxide surface layers show an O1s peak in the typical oxide region between 529-530eV, even more, the values obtained are ~1eV shifted from values found in the literature. The O1s peak is relatively broad and symmetric; its deconvolution shows two components for the reference foil as well as for the plasma-treated Al foil with BEs of 532.21 and 533.45 eV. In the case of plasma-treated Al foil, a clear difference in the ratios of these two components can be seen. The peak at 532.21 eV BE related to $Al_2O_3$ increases after plasma treatment, indicating an increase of the oxide thickness. This assumption can easily be corroborated using the calculations of the type developed by Strohmeier,[19] yielding a small increase of the oxide thickness from 4.6 to 4.8 nm thickness. From the O1s spectra in Figure 5, it can also be seen that there is a decrease of the carbon-oxygen C=O-related line at 533.45 eV BE.

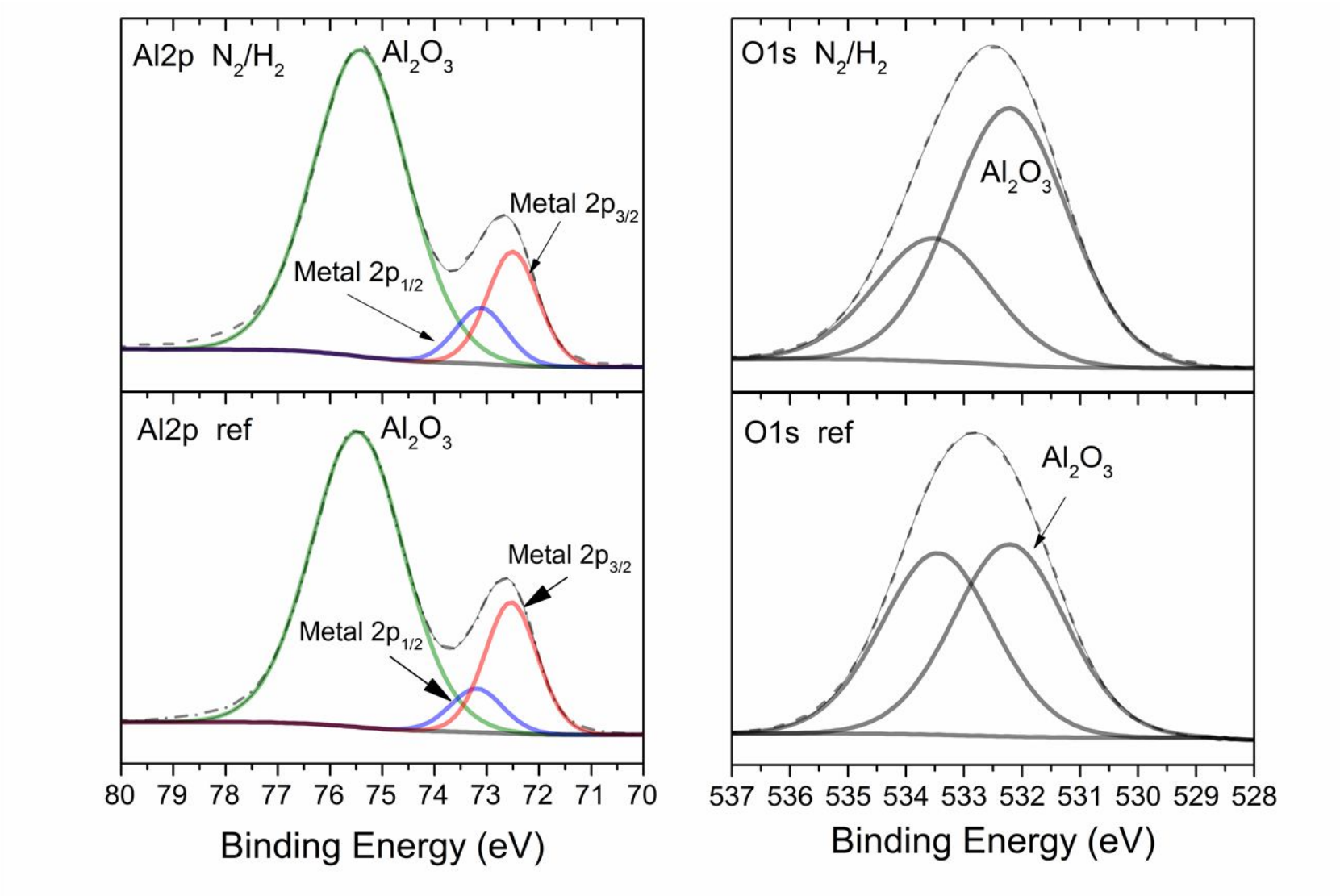

Figure 5: Al 2p (left column) and O 1s (right column) XPS core level spectra of Al metal foils. Bottom row: Pristine Al reference sample; top row: Al foil after a-C contamination and subsequent $N_2/H_2$ plasma treatment.

### *3.1 Al EUV Filter Cleaning*

Taking into account the above results regarding the performance of the $N_2/H_2$ plasma and especially its interesting feature on the preservation of the original surface chemistry of the plasma-processed metal foils, one can conclude that this process would be sufficiently safe for the carbon contamination cleaning of the corresponding thin Al filter foils. Here, besides restoring the original optical performance of the pure Al metal filter material the preservation of the mechanical integrity of these devices is of prime importance. Thus, avoiding an increase of the native $Al_2O_3$ surface layer thickness is of importance for both the optical as well as the mechanical performance of these self-sustained ultrathin Al filter foils.

We have thus performed a $N_2/H_2$ plasma cleaning on a set of Al EUV foil filters that have been exposed for roughly 500 operation hours to the pulsed high brilliance photon beam of the FERMI FEL1 in a photon energy range of 20 to 65 eV. The results from the above plasma cleaning process are shown in Figure 6. As can be seen from there, a successful carbon removal was obtained on the considerably more carbon-contaminated filter upstream side. An additional interesting feature consists of the fact that the carbon-contaminated Al filter foils exhibit a wrinkled appearance at the position of the carbon footprint (Figure 6A). On the other hand, as can be seen from Figure 6B after the plasma cleaning, the cleaned filter surfaces do not show these wrinkles any further. We conclude from this that the foil wrinkles do result from the fact that the carbon layer is deposited onto the Al metal at elevated temperatures during the photon beam exposure. After cooling down, the difference in linear thermal coefficients between these two materials (aluminum: 21 to 25 x $10^{-6}$ m (m K)$^{-1}$; graphite: 4 to 8 x $10^{-6}$ m (m K)$^{-1}$; RT values) results in the observed formation of wrinkles associated with mechanical stress at the carbon/Al interface. Thus, the removal of the carbon footprint simultaneously does remove the observed wrinkles within the Al filter foils, becoming stress-free again.

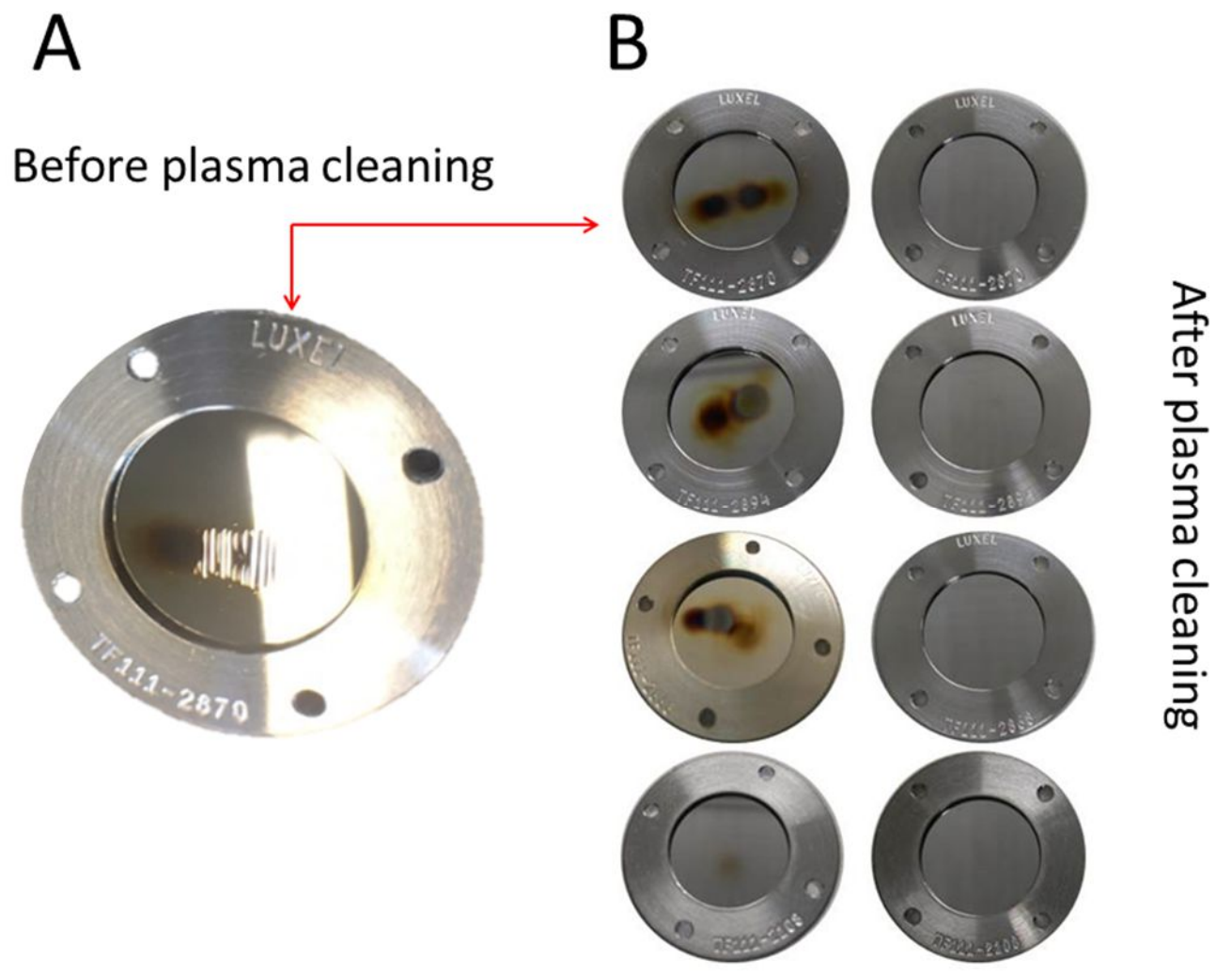

Figure 6: **A.** Upstream side of a self-sustained Al filter foil with 100 nm thickness used at the FERMI FEL1. The foil wrinkles within the carbon footprint induced by the deposition of the carbon layer at elevated foil temperatures can be clearly distinguished. **B.** Upstream side of four different self-sustained Al filter foils with 100 nm foil thickness used at the FERMI FEL1 before (left column) and after (right column) $N_2/H_2$ plasma treatment. Several photon beam-induced carbon footprints per each filter foil can clearly be distinguished from the images in the left-hand side column

### *3.1 Carbon Contamination Characterization and Plasma Cleaning Mechanism*

Raman spectroscopy was used to characterize the different carbon contaminations on the various samples in order to establish a comparison between the carbon cleaning characteristic of the carbon deposited on the EUV filters and the carbon deposited on the metal foils. Figure 7 shows the Raman spectra for the different carbon contaminations, plus a carbon contamination from a gold-coated mirror used at a soft-ray bending magnet beamline at the former Synchrotron Radiation Center (SRC). The most intense contamination corresponds to the upstream side of the EUV filter as can be seen from Figure 7, showing two characteristic carbon-related peaks denoted as D (1378 $cm^{-1}$) and G (1542 $cm^{-1}$). Although these peaks can be distinguished, the carbon content is related to amorphous carbon (a-C) with little graphitic ordering, the same accounts for the contamination from the back side and the SRC carbon contamination.

Note that in graphite the G mode is at 1581 $cm^{-1}$ involving the in-plain bond-stretching motion of C $sp^2$ atom pairs, while the D peak lies around 1355$cm^{-1}$ and is known as a defect mode. This mode is forbidden in perfect graphite and only gets active in presence of disorder[20]. In this context, it is possible to deduce that the carbon contaminations on the Al filter foils also do not show a purely graphitic behavior, as they rather consist of a mixture of $sp^2$ and $sp^3$ hybridized carbon consisting of broad and smooth Raman spectral features. Hence, the broadening the Raman bands and the high intensity of the D peak relative to the G peak indicate a low graphitic content in the carbon contamination footprint[21]. Summarizing, there is a somewhat surprising overall chemical and structural similarity between the a-C deposits from both the SRC bending magnet source and the FERMI FEL1 free electron laser source but with a variance from the pure amorphous character of the carbon layer as generated by the e-beam deposition process.

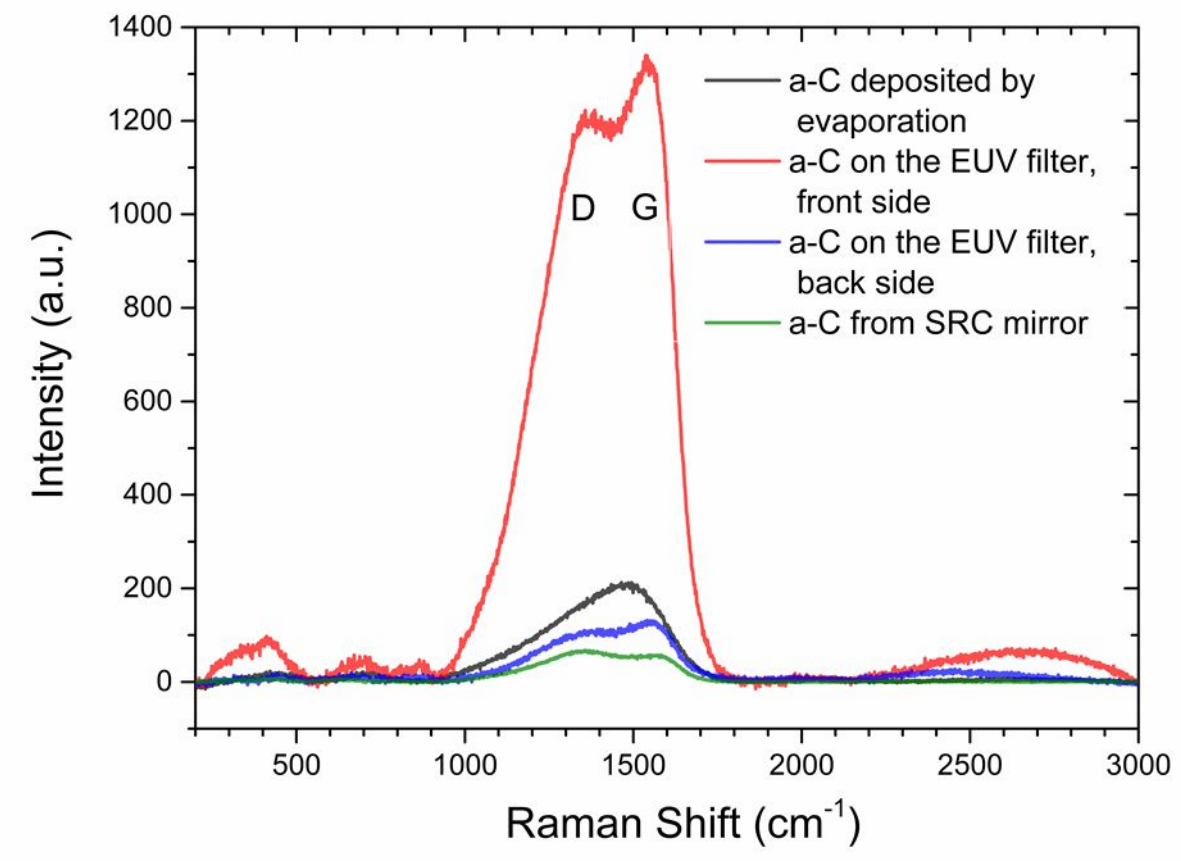

Figure 7: Raman spectra from different carbon contaminations. Red solid line: Carbon footprint on EUV Al filter foil upstream side; Black solid line: a-C thin film from e-beam carbon deposition; Blue solid line: Carbon footprint on EUV Al filter foil downstream side; Green solid line: Carbon footprint an Au-coated from SRC soft X-ray beamline.

By using the differentially pumped RGA within the experimental setup (see Figure 2), we could confirm the formation of $NH_3$ species within the $N_2/H_2$ plasma. Figure 8 shows the RGA time trend from that $N_2/H_2$ plasma cleaning process of the EUV Al foil filters. Previous studies describe ammonia formation in this type of plasma either by the adsorption of excited $N_2$ molecules and $N_2^+$ ions, followed by their dissociation at the stainless steel chamber inside walls and then recombining with atomic hydrogen from the gas phase or, alternatively, by the direct absorption of atomic N and H plasma species instead of dissociative adsorption[22].

Therefore, we can attribute the cleaning process mainly to the ammonia, arguing that $NH_3$ will reduce the carbon within the contamination layer, the cleaning rate thereby not only depend on the total pressure (via mean the corresponding free path lengths) but obviously also on the ammonia production rate/concentration within the plasma. Decomposition of the hydrocarbon molecules takes place by breaking the C-H strong chemical bonds.

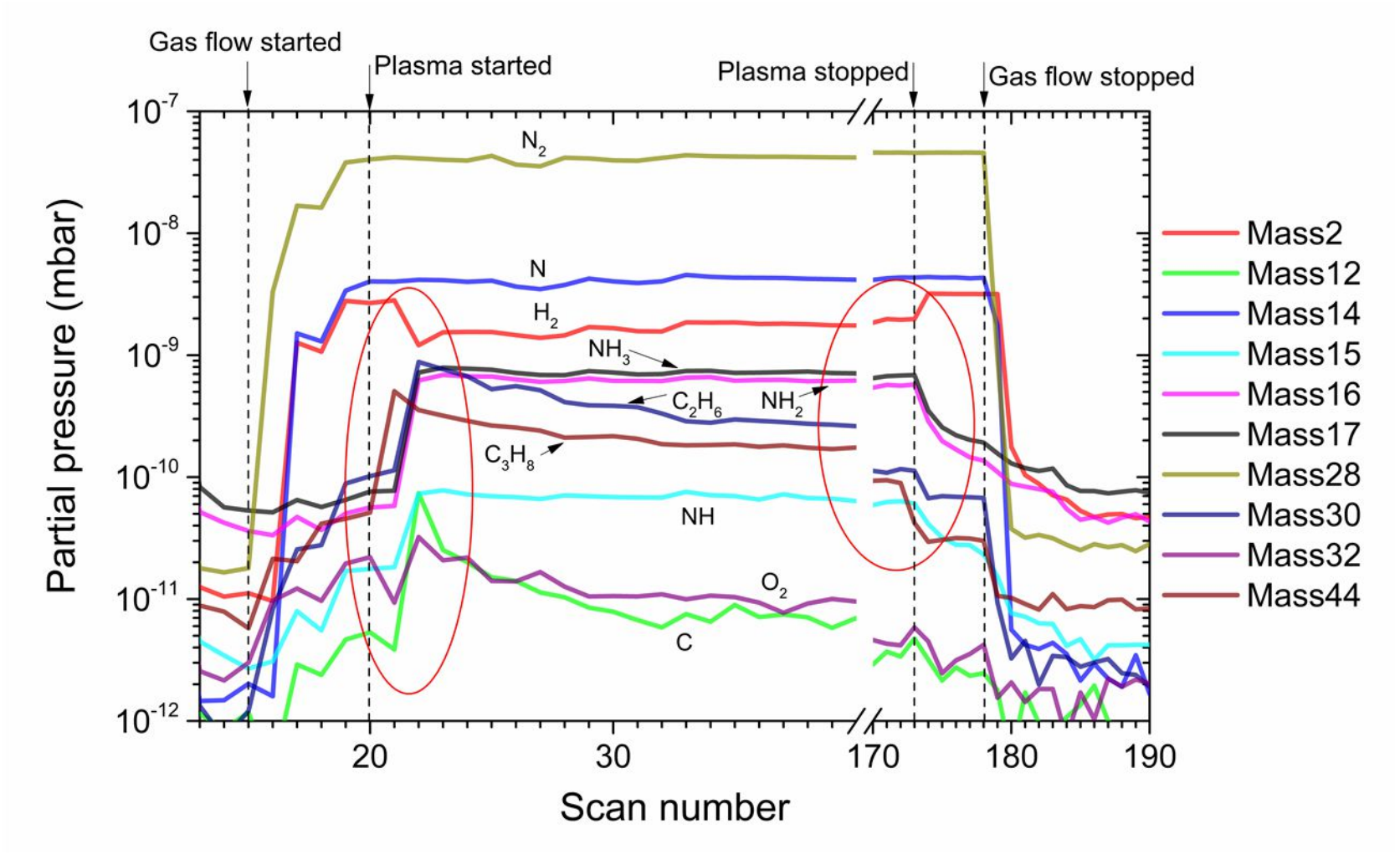

Figure 8: RGA time trend from the $N_2/H_2$ plasma during the plasma cleaning of the EUV Al foil filters. The complete time span of the RGA trend plot corresponds to about 4 hours and 50 minutes.

According to the above and as can be seen from Figure 8, the ignition of the plasma (at the time of scan 20) results in the formation of $NH_3$ species together with corresponding $NH_2$ and NH cracking products that also vanish once the plasma is stopped (at the time of scan 173). The gaseous reaction products from the interaction between the carbon and the ammonia species equally do show the same dependence on the plasma operation, as the ethane (mass 30) and propane (mass 44) mass trends appear and disappear with the plasma being on or off, respectively (see the read markups in Figure 8). From the decrease of these two time trends during while the plasma is operative, one can conclude on the depletion of carbon species within the plasma chamber as a function of the plasma operation time.

Table I: XPS core level binding energies as obtained from the XPS spectra presented in Figs. 3 to 5.

| **XPS line** | **Reference foil<br>B. E. (eV)** | **$N_2/O_2/H_2$ Plasma<br>B. E. (eV)** | **$N_2/H_2$ Plasma<br>B. E. (eV)** |
|---|---|---|---|
| **Ni metal foil** | | | |
| Ni 2p | 852.08 Metal | 852.1 Metal | 852.08 Metal |
| | 855.36 $Ni_2O_3$ | 853.58 NiO | 855.41 $Ni_2O_3$ |
| | | 855.71 NiO (+ NiOOH) | |
| O 1s | 529.02 $Ni_2O_3$ | 529.06 NiO | 529.01 Ni2O3 |
| | 530.84 C-C | 530.68 C-C | 530.93 C-C |
| | 531.67 O-H, O with C | 531.68 O-H, O with C | 531.88 O-H, O with C |
| | 532.93 | 532.68 | 533.32 Nitrates |
| **Rh metal foil** | | | |
| Rh 3d | 307.04 Metal | 307.06 Metal | 307.05 Metal |
| | 307.60 $Rh_2O_3$ | 307.93 $Rh_2O_3$ | 308.27 $Rh_2O_3$ |
| | | 309.59 RhOOH | |
| O 1s | 529.52 $Rh_2O_3$ | 529.68 $Rh_2O_3$ | 530.23 $Rh_2O_3$ |
| | 531.59 Carbonates | 531.32 RhOOH | 532.07 Carbonates |
| | | 532.49 Carbonates | |
| | | 534.35 Absorbed $H_2O$ | |
| **Al metal foil** | | | |
| Al 2p | 72.53 Metal $2p_{3/2}$ | | 72.5 Metal $2p_{3/2}$ |
| | 73.19 Metal $2p_{1/2}$ | | 73.1 Metal $2p_{1/2}$ |
| | 75.47 $Al_2O_3$ | | 75.41 $Al_2O_3$ |
| O 1s | 532.21 $Al_2O_3$ | | 532.21 $Al_2O_3$ |
| | 533.45 C=O | | 533.52 C=O |

## 4 CONCLUSIONS

Summarizing, we conclude that a favorable combination of a $N_2$(95%)/$H_2$(5%) plasma feedstock gas mixture leads to the best chemical surface preservation results for non-inert metal optical coatings such as Ni, Rh, and Al together with the removal of carbon contaminations and including an acceptable carbon cleaning rate. However, the above feedstock gas mixture does not remove carbon contaminations as fast as, e.g., a $N_2$(94%)/$O_2$(4%)/$H_2$(2%) plasma which, on the other hand, induces the surface formation of NiO and NiOOH in Ni and RhOOH in Rh foils. Thus, the application of a specific gas mixture for low-pressure RF plasma has to be considered on the background of the beamline optics requirements. It should be emphasized in this context that the above metal surface preservation has been demonstrated for the specific cases of Rh, Ni, and Al and cannot be necessarily be transferred directly to other metals without prior testing.

As an application of the above $N_2$(95%)/$H_2$(5%) plasma, we could demonstrate a first efficient ex-situ carbon cleaning process of self-sustained thin Al foil optical filters (100 nm thickness) as a specific case study within the field of transmission optics for FEL application in the EUV photon energy range. As a peculiar side effect, macroscopic wrinkles in the Al foil located within the carbon footprint were removed simultaneously.

**AUTHOR INFORMATION**

*E-mail: epellegrin@cells.es. Phone +34 935924418.

**ORCID**

H. Moreno Fernández: 0000-0002-4362-9488

**Author contributions**

**Notes**

The authors declare no competing financial interest.

## ACKNOWLEDGEMENTS

The research by HMF is supported by funding from the "Generalitat de Catalunya, Departament d'Empresa i Coneixement" within the "Doctorats Industrials" program (dossier no. 2014 DI 037). In addition we acknowledge the expert assistance by M. Dominguez Escalante (UPC, Barcelona) for the XPS measurements as well as by F. J. Belarre Triviño (ICN2, Bellatera) for the carbon deposition process.